\documentstyle[12pt]{article}

\def\RR {\bf R}
\def\CC {\bf C}

\newcommand{\vs}[1]{\vspace{#1 mm}}

\begin{document}
\begin{titlepage}

\setcounter{page}{0}
\begin{flushright}
KEK Preprint 98-99 \\
OU-HET 297 \\
hep-th/9807111
\end{flushright}

\vs{5}
\begin{center}
{\Large\bf More on the Similarity between
$D=5$ Simple Supergravity and M Theory}
\vs{15}

{\large
Shun'ya Mizoguchi\footnote{mizoguch@tanashi.kek.jp}} \\
\vs{5}
{\em Institute of Particle and Nuclear Studies \\
High Energy Accelerator Research Organization, KEK \\
Tanashi, Tokyo 188-8501, Japan} \\
\vs{5}
and \\
\vs{5}
{\large
Nobuyoshi Ohta\footnote{ohta@phys.wani.osaka-u.ac.jp}
} \\
\vs{5}
{\it Department of Physics, Osaka University,
Toyonaka, Osaka 560-0043, Japan}
\end{center}
\vs{10}
\centerline{{\bf{Abstract}}}
\vskip 3mm
It has been known that $D=5$ simple supergravity resembles $D=11$
supergravity in many respects. We present their further resemblances in
(1) the duality groups upon dimensional reduction, and (2) the worldsheet
structure of the solitonic string of the $D=5$ supergravity. We show that
the $D=3$, $G_{2(+2)}/SO(4)$ (bosonic) nonlinear sigma model is obtained
by using Freudenthal's construction in parallel to the derivation of the
$D=3$, $E_{8(+8)}/SO(16)$ sigma model from $D=11$ supergravity. The zero
modes of the string solution with unbroken (4,0) supersymmetry consist of
three (non-chiral) scalars, four Majorana-Weyl spinors of the same
chirality and one chiral scalar, which suggests a duality to a certain
six-dimensional chiral string theory. The worldsheet gravitational anomaly
indicates a quantum correction to the Bianchi identity for the dualized
two-form gauge field in the bulk just like the M5-brane case.

\end{titlepage}

\baselineskip=18pt
\setcounter{footnote}{0}
\def\beq{\begin{equation}}
\def\eeq{\end{equation}}
\def\beqa{\begin{eqnarray}}
\def\eeqa{\end{eqnarray}}
\def\n{\nonumber\\}
\def\ld{\lambda}
\def\sm{\sigma}

\def\mb{\overline{\mu}}
\def\nb{\overline{\nu}}
\def\ab{\overline{\alpha}}
\def\beb{\overline{\beta}}

\def\ep{e_{\dot{+}}^+}
\def\emi{e_{\dot{-}}^-}
\def\pld{\dot{+}}
\def\mnd{\dot{-}}

\def\ad{\mbox{ad}}

\section{Introduction}

It has been known for a long time that $D=5$ simple supergravity
resembles $D=11$ supergravity~\cite{Cr,ChNi}. The name of ``simple''
supergravity is given because it contains a possible minimal set of
fields (a single $N=2$ gravity multiplet). It consists of one graviton,
two $N=2$ symplectic-Majorana gravitini which are equivalent to a
single Dirac gravitino, and one $U(1)$ gauge field which replaces
the three-form gauge field in $D=11$ supergravity~\cite{CrJuSc}.
The Lagrangians are exactly in the same form~\cite{ChNi} except for the
difference in the number of indices of the gauge fields. It is also
known that the dimensional reduction to $D=4$~\cite{CrJu} can be carried
out in a similar way.

In this paper we will present their further resemblances in two
different aspects. In the first part we study the dimensional reduction
to three dimensions and show that the $D=3$, $G_{2(+2)}/SO(4)$ (bosonic)
nonlinear sigma model can be obtained by using Freudenthal's construction
of the exceptional Lie algebras. The derivation is completely parallel to
that of the $D=3$, $E_{8(+8)}/SO(16)$ sigma model from $D=11$
supergravity~\cite{E10}. Although the appearance of $G_{2(+2)}/SO(4)$ in
this reduction is already known by the r- and c-maps
($G_{2(+2)}/SO(4)$ is the image of the empty space as a very special
real manifold. See \cite{c-map} for a review.), the explicit construction
makes the similarity of the two supergravity theories transparent. In
particular, Freudenthal's construction reveals how similar the structures
of $E_{8(+8)}/SO(16)$ and $G_{2(+2)}/SO(4)$ are.

In the second part we investigate the worldsheet structure of the
magnetically charged BPS solitonic string of the $D=5$ supergravity.
As we shall see, this string has a number of properties which are similar 
to those of M5-brane. For example, unbroken worldsheet supersymmetry 
is chiral, and it also has a chiral bosonic massless excitation which may 
be viewed as an analogue of the self-dual two-form on M5-brane. Two 
is again the number of world-volume dimensions where a gravitational 
anomaly may exist, and indeed it does. This (tangent-bundle) anomaly in 
turn indicates a quantum correction to the Bianchi identity for the 
dualized two-form gauge field in the bulk just like the M5-brane 
case~\cite{DLM,VW}. We will not discuss possible normal-bundle 
anomaly~\cite{Wi,FHMM} in this paper.

Of course, $D=5$ simple 
supergravity can be realized as a Calabi-Yau compactification of 
$D=11$ supergravity~\cite{CCDAF,Fe} with $h_{11}=1$, together with the 
truncation of scalar multiplets. (This truncation is necessary since 
there arises at least one scalar multiplet for any Calabi-Yau 
compactification.) Therefore, its properties can be explained in terms 
of the $D=11$ theory. For example, the string in the $D=5$ supergravity 
can be viewed as M5-brane wrapped around the unique 4-cycle of this 
Calabi-Yau, while the 0-brane dual to the string is M2-brane wrapped 
around the unique 2-cycle dual to the 4-cycle. We should emphasize here 
that our aim is not to investigate the properties of this particular kind 
of supergravity itself. Rather, we would like to convince the reader that 
this $D=5$ model may afford a nice testing ground for various ideas in 
$D=11$ supergravity and M theory, hoping that we could get an insight 
into the unsolved interesting problems of these theories in a simpler 
setting.

\section{Dimensional reduction and nonlinear sigma models}
\subsection{Reduction to $D=4$}
We first briefly review the dimensional reduction of simple $D=5$ 
supergravity to $D=4$~\cite{ChNi}. We present it in such a way that the 
$SL(2,{\bf R})$ symmetry may look similar to the $E_{7(+7)}$ symmetry of 
$N=8$ supergravity. We also identify the discrete quantum duality 
group~\cite{Uduality}.

Retaining only the bosonic degrees of freedom, the Lagrangian of simple 
$D=5$ supergravity is given by
\beq
{\cal L}
=E^{(5)}(R^{(5)}-\frac14 F_{MN}F^{MN})
-\frac1{12\sqrt{3}}\epsilon^{MNPQR}F_{MN}F_{PQ}A_R ,
\label{5DL}
\eeq
where $F_{MN}=2\partial_{[M}A_{N]}$. We take the signature $(-++++)$. 
Dropping the fifth-coordinate dependence, 
we get the four-dimensional Lagrangian
\beqa
{\cal L}
&=& E^{(4)}\left(R^{(4)}
-\frac32 \partial_{\hat{\mu}}\ln\rho \partial^{\hat{\mu}}\ln\rho
-\frac12\rho^{-2} \partial_{\hat{\mu}}A \partial^{\hat{\mu}}A
-\frac14\rho^3 B_{\hat{\mu}\hat{\nu}} B^{\hat{\mu}\hat{\nu}}
\right.\n
&&\left.
-\frac14\rho F^{(4)}_{\hat{\mu}\hat{\nu}}
F^{(4)\hat{\mu}\hat{\nu}}
-\frac1{4\sqrt{3}}E^{(4)-1}
\epsilon^{\hat{\mu}\hat{\nu}\hat{\rho}\hat{\sigma}}
F_{\hat{\mu}\hat{\nu}}F_{\hat{\rho}\hat{\sigma}}A
\right),
\label{4DL}
\eeqa
where the f\"unfbein and the vector field are parameterized as
\beq
E_{~M}^{(5)A}=\left[
\begin{array}{cc}\rho^{-\frac12}E_{~\hat{\mu}}^{(4)\hat{\alpha}}
& \rho B_{\hat{\mu}} \\
0&\rho\end{array}
\right], \;
A_M = [A_{\hat{\mu}}, A].
\label{f}
\eeq
$\hat{\mu},\hat{\nu},\ldots$ and $\hat{\alpha},\hat{\beta},\ldots$
are four-dimensional curved and flat indices, respectively.
$F^{(4)}_{\hat{\mu}\hat{\nu}}$ is defined by 
$F^{(4)}_{\hat{\mu}\hat{\nu}}
 \equiv F'_{\hat{\mu}\hat{\nu}} + B_{\hat{\mu}\hat{\nu}}A$,
where  
$F'_{\hat{\mu}\hat{\nu}}
=2\partial_{[\hat{\mu}}A'_{\hat{\nu}]}$ is the field strength 
of the Kaluza-Klein invariant vector field 
$A'_{\hat{\mu}}=A_{\hat{\mu}}-B_{\hat{\mu}} A$, and  
$B_{\hat{\mu}\hat{\nu}} = 2 \partial_{[{\hat{\mu}}} B_{\hat{\nu}]}$.
We dualize $A'_{\hat{\mu}}$ field by adding
\beq
{\cal L}_{\rm Lag.mult.}=\frac12 
\epsilon^{\hat{\mu}\hat{\nu}\hat{\rho}\hat{\sigma}}
\tilde{A}_{\hat{\sigma}}\partial_{\hat{\rho}}F'_{\hat{\mu}\hat{\nu}}
\label{LLm}
\eeq
to ${\cal L}$. 
Up to an irrelevant perfect square  
one finds 
\beqa
{\cal L}+{\cal L}_{\rm Lag.mult.}
&=&E^{(4)}R^{(4)} + {\cal L}_S +{\cal L}_V,\n
{\cal L}_S
&=&-E^{(4)}(\frac32\partial_{\hat{\mu}}\ln \rho
                        \partial^{\hat{\mu}}\ln \rho
                 +\frac12\rho^{-2}\partial_{\hat{\mu}}A
                        \partial^{\hat{\mu}}A),\n
{\cal L}_V
&=&-\frac14E^{(4)}{\cal G}^T_{\hat{\mu}\hat{\nu}}
N^{\hat{\mu}\hat{\nu}\hat{\rho}\hat{\sigma}}
{\cal G}_{\hat{\rho}\hat{\sigma}},
\label{4d}
\eeqa
where ${\cal G}_{\hat{\mu}\hat{\nu}}
=\left[\begin{array}{c}\tilde{A}_{\hat{\mu}\hat{\nu}}\\
B_{\hat{\mu}\hat{\nu}}\end{array} \right]$, 
${\tilde A}_{\hat{\mu}\hat{\nu}}= 2 \partial_{[{\hat{\mu}}}
{\tilde A}_{\hat{\nu}]}$. 
$N^{\hat{\mu}\hat{\nu}\hat{\rho}\hat{\sigma}}$ is a two-by-two matrix 
and can be written as
\beqa
N^{\hat{\mu}\hat{\nu}\hat{\rho}\hat{\sigma}}
&=& m~{\bf 1}^{\hat{\mu}\hat{\nu}\hat{\rho}\hat{\sigma}}
+a~(\star)^{\hat{\mu}\hat{\nu}\hat{\rho}\hat{\sigma}},\n
V^{-1}mV^{-1}&=&K-\frac12(\Phi\Phi^*K+K\Phi^*\Phi)
+\frac14\Phi\Phi^{*2}K\Phi,\n
V^{-1}aV^{-1}&=&-\Phi^*K-\Phi+\frac12(\Phi\Phi^{*2}K+K\Phi^{*2}\Phi)
 +\frac13\Phi\Phi^*\Phi-\frac14\Phi\Phi^{*3}K\Phi,
\eeqa
where 
${\bf 1}^{\hat{\mu}\hat{\nu}\hat{\rho}\hat{\sigma}}
= \frac12(G^{(4)\hat{\mu}\hat{\rho}}G^{(4)\hat{\nu}\hat{\sigma}}
-G^{(4)\hat{\nu}\hat{\rho}}G^{(4)\hat{\mu}\hat{\sigma}})$, 
$(\star)^{\hat{\mu}\hat{\nu}\hat{\rho}\hat{\sigma}}
=\frac12 E^{(4)-1}\epsilon^{\hat{\mu}\hat{\nu}\hat{\rho}\hat{\sigma}}$, 
$V = \left[\begin{array}{cc}\rho^{-\frac12}& 0 \\
 0 &\rho^{\frac32}\end{array}
\right]$, 
$\Phi=\left[\begin{array}{cc}
 0 &\sqrt{3}\phi\\
 \sqrt{3}\phi& 0 \end{array}
\right]$, \
$\Phi^*=\left[\begin{array}{cc} 2\phi& 0 \\ 0 &0\end{array}
\right]$,
$K =(1+\Phi^{*2})^{-1}$ and 
$\phi=\frac1{\sqrt{3}}\rho^{-1}A$.

To see the duality symmetry we define
${\cal H}_{\hat{\mu}\hat{\nu}}$ by the equation 
\beq
{\cal H}_{\hat{\mu}\hat{\nu}}
=m (\star {\cal G})_{\hat{\mu}\hat{\nu}}-a {\cal G}_{\hat{\mu}\hat{\nu}}.
\eeq
It can be shown that 
${\cal G}_{\hat{\mu}\hat{\nu}}$ and ${\cal H}_{\hat{\mu}\hat{\nu}}$
are related by
\beq
{\cal F}_{\hat{\mu}\hat{\nu}} \equiv
\left[\begin{array}{c}
{\cal G}_{\hat{\mu}\hat{\nu}} \\ {\cal H}_{\hat{\mu}\hat{\nu}}
\end{array}\right]
=\Omega{\cal V}^{(4)T}{\cal V}^{(4)}
\left[\begin{array}{c}
\star {\cal G}_{\hat{\mu}\hat{\nu}} \\ 
\star {\cal H}_{\hat{\mu}\hat{\nu}}
\end{array}\right],
\eeq
where $\Omega$ and ${\cal V}^{(4)}$ are four-by-four matrices defined by
$\Omega=\left[\begin{array}{cc}&-{\bf 1}\\{\bf 1}&\end{array}\right]$, 
${\cal V}^{(4)}={\cal V}^{(4)}_-{\cal V}^{(4)}_+$,
${\cal V}^{(4)}_-=\exp\left[\begin{array}{cc}&-\Phi^*\\-\Phi&
\end{array}\right]$
and ${\cal V}^{(4)}_+=\left[\begin{array}{cc}V&\\&V^{-1}
\end{array}\right]$.
Writing
\footnote{Note that ${\cal V}^{(4)}={\cal V}^{(4)}_+{\cal V}^{(4)c}_-$, 
where ${\cal V}^{(4)c}_-={\cal V}^{(4)-1}_+
{\cal V}^{(4)}_-{\cal V}^{(4)}_+ =P^{-1}\exp(-\frac1{\sqrt{3}}AE)P$.
} 
\beqa
{\cal V}^{(4)}_-=P^{-1}\exp(-\phi E)P,~~
{\cal V}^{(4)}_+=P^{-1}\exp(\ln \rho^{-\frac12} H)P,
\nonumber
\eeqa
\beqa \!\!\!
P = \left[\begin{array}{cccc}
0&0&0&1\\
1&0&0&0\\
0&0&1&0\\
0&1&0&0
\end{array}\right]=(P^{-1})^T,~~ 
E=\left[\begin{array}{cccc}
0&\sqrt{3}&&\\
&0&2&\\
&&0&\sqrt{3}\\
&&&0
\end{array}\right],~~  
H = \left[\begin{array}{cccc}
3&&&\\
&1&&\\
&&-1&\\
&&&-3
\end{array}\right],
\eeqa  
it is clear that  $\phi$ and $\ln \rho$ parameterize 
$SL(2,{\bf R})/SO(2)$. Note that $H,E$ and $F\equiv E^T$
are the ${\bf 4}$ representation matrices of the Chevalley generators 
of the Lie algebra of $SL(2,{\bf R})$ embedded in $Sp(4)$ 
(defined with respect to $P\Omega P^{-1}$). 
Denoting ${\cal R}={\cal V}^{(4)T}{\cal V}^{(4)}$, one may write 
\beq
{\cal L}_V+{\cal L}_S
= E^{(4)} \left(\frac14 {\cal G}^T_{\hat{\mu}\hat{\nu}}
\star {\cal H}^{\hat{\mu}\hat{\nu}}
+ \frac3{40}{\rm Tr}\partial_{\hat{\mu}}{\cal R}^{-1}
\partial^{\hat{\mu}}{\cal R}\right).
\eeq
The field equations are invariant under a rigid transformation
${\cal F}_{\hat{\mu}\hat{\nu}}\to\Lambda^{-1}{\cal F}_{\hat{\mu}\hat{\nu}}$,
${\cal R}\rightarrow \Lambda^T{\cal R}\Lambda$ for $\Lambda\in SL(2,{\bf R})$.
The transformation of ${\cal V}^{(4)}$ must be accompanied by a compensating
local $SO(2)$ transformation ${\cal V}^{(4)}\rightarrow h(x){\cal V}
^{(4)}\Lambda$,
$h(x)\in \{
P^{-1}\exp(t(E-F)) P~|~t\in {\bf R}
\}$, as usual, so as to restore the parameterization
of the coset space. 

Typically, such a continuous symmetry of a classical supergravity is
broken at the quantum level to a discrete subgroup due to the charge
quantization condition. In our case the classical duality group 
$SL(2,{\bf R})$ is broken to the intersection $SL(2,{\bf R})\cap 
Sp(4,{\bf Z})$, where the latter is defined as a discrete subgroup of 
$Sp(4)$ that preserves the charge lattice 
and must be represented by matrices with integer entries in some basis.
To identify this we change the charge normalization 
as ${\cal F}'=U{\cal F}$, ${{\cal V}^{(4)}}'=U{\cal V}^{(4)}U^{-1}$, etc.  
by the constant matrix $U=\left[\begin{array}{cccc}
1&&&\\&\sqrt{3}&&\\&&1&\\&&&\sqrt{3}\end{array}
\right]$ (so that the local $SO(2)$ is now
defined as the group of matrices that preserves the `metric' $U^{-2}$)
\footnote{The normalization of $\tilde{A}_{\hat{\sigma}}$ 
is so chosen that the matrix $\Phi$ 
becomes symmetric, and therefore the involutive 
automorphism is obtained by simply transposing the inverse. If we had 
taken $\sqrt{3}/2$ instead of $1/2$ as the factor in (\ref{LLm}), we 
could have directly derived $E$,$F$ and $H$ with integer entries, but 
as the price the involutive automorphism then becomes less trivial. 
}. 
Then the generators of the Lie algebra of $SL(2,{\bf R})$ are transformed 
into
$E'=\left[\begin{array}{cccc}0&3&&\\&0&2&\\&&0&1\\&&&0\end{array}\right]$,
$F'=\left[\begin{array}{cccc}0&&&\\1&0&&\\&2&0&\\&&3&0\end{array}\right]$
and $H'=H$. In this basis
$SL(2,{\bf Z})\equiv SL(2,{\bf R})\cap Sp(4,{\bf Z})$ is realized as
matrices with integer entries. The whole duality group is generated by
the modular group generators $P^{-1}SP$ and $P^{-1}TP$,
where
\beq
S= \exp(-F') \exp E' \exp(-F')
=\left[\begin{array}{cccc}&&&1\\&&-1&\\&1&&\\-1&&&\end{array}\right],
~~T= \exp E'
=\left[\begin{array}{cccc}
1&3&3&1\\&1&2&1\\&&1&1\\&&&1\end{array}\right].
\eeq
Note that on the scalar $Z\equiv -\phi+i\rho$ this definition of 
$SL(2,{\bf Z})$ induces the familiar modular transformations $Z \to Z+1$
and $Z \to -1/Z$ under $T$ and $S$, respectively.

\subsection{Reduction to $D=3$}
We will now turn to the reduction to $D=3$. 
Splitting the curved indices $M,N,\ldots=0,\ldots,4$ into
$\mu,\nu,\ldots=0,1,2$ and $i,j,\ldots=3,4$ and ignoring the dependence 
of the latter coordinates, the original Lagrangian~(\ref{5DL}) is reduced 
to
\beqa
{\cal L}&=& E^{(3)}\left(R^{(3)}
-\partial_\mu\ln e\partial^\mu\ln e
+\frac14 \partial_\mu g_{ij} \partial^\mu g^{ij}\right.\n
&&
-\frac14 e^{-2}g_{ij} B_{\mu\nu}^i B^{j\mu\nu} 
-\frac14 e^2 F^{(3)}_{\mu\nu}F^{(3)\mu\nu}
-\frac12 g^{ij}\partial_\mu A_i \partial^\mu A_j\n
&&\left.-\frac1{2\sqrt3}E^{(3)-1}\epsilon^{\mu\nu\rho}\epsilon^{ij}
 F_{\mu\nu}\partial_{\rho}A_i\cdot A_j \right),
\label{3DL}
\eeqa
where the f\"{u}nfbein and the vector field are decomposed as 
\beq
E_{~M}^{(5)A}=\left[
\begin{array}{cc}e^{-1}E_{~\mu}^{(3)\alpha}&B_\mu^ie_i^{~a}\\
0&e_i^{~a}\end{array}
\right], \;
A_M = [A_{\mu}, A_i].
\eeq
The flat Lorentz indices $A,B,\ldots =0,\ldots,4$ are split into
$\alpha,\beta,\ldots=0,1,2$ and $a,b,\ldots=3,4$.
In eq.~(\ref{3DL}) (as well as in eq.~(\ref{3DL1}) below),
$\mu,\nu,\ldots$ are raised by the metric $G^{(3)\mu\nu}\equiv
E_{~\alpha}^{(3)\mu}E_{~\beta}^{(3)\nu}\eta_{\alpha\beta}$,
$\eta_{\alpha\beta}=\mbox{diag}[-1,+1,+1]$;
$E^{(5)}$, $E^{(3)}$ and $e$ are the determinants of $E_{~M}^{(5)A}$,
$E_{~\mu}^{(3)\alpha}$ and $e_i^{~a}$, respectively;
$B_{\mu\nu}^i=2\partial_{[\mu}B_{\nu]}^i$ is the Kaluza-Klein gauge
field strength. As in $D=4$ case, $F^{(3)\mu\nu}$ is defined by the 
equation $F^{(3)}_{\mu\nu}\equiv F'_{(D=3)\mu\nu}+B_{\mu\nu}^iA_i$,
where $F'_{(D=3)\mu\nu}$ is the field strength of the Kaluza-Klein
invariant vector field $A'_{(D=3)\mu}=A_\mu-B_\mu^iA_i$.

These three-dimensional vector fields are dualized by introducing
the Lagrange multiplier fields
\beq
{\cal L}_{\mbox{\scriptsize Lag.mult.}}=
\frac12\epsilon^{\mu\nu\rho}(\varphi\partial_{\mu}F'_{(D=3)\nu\rho}
+\psi_i\partial_{\mu}B^i_{\nu\rho}).
\label{LLag}
\eeq
Up to irrelevant perfect squares of $A'_{(D=3)\mu\nu}$ and $B^i_{\mu\nu}$,
one finds 
\beqa
{\cal L}+{\cal L}_{\mbox{\scriptsize Lag.mult.}}
&=&
E^{(3)}\left(R^{(3)}
-\partial_\mu\ln e\partial^\mu\ln e
+\frac14 \partial_\mu g_{ij} \partial^\mu g^{ij}
-\frac12 g^{ij}\partial_\mu A_i \partial^\mu A_j\right.\n
&&
-\frac12 e^{-2}(\partial_\mu\varphi
 -\frac{1}{\sqrt{3}}\epsilon^{ij}A_i\partial_\mu A_j)
  (\partial^\mu\varphi
 -\frac{1}{\sqrt{3}}\epsilon^{kl}A_k\partial^\mu A_l)
\n
&&
-\frac12e^{-2}g^{ij}
  (\partial_\mu\psi_i
   +\frac12(\varphi\partial_\mu A_i-A_i\partial_\mu\varphi)
 +\frac{1}{3\sqrt{3}}\epsilon^{k_1l_1}A_iA_{k_1}\partial_\mu A_{l_1})
\n
&&~~~~~\cdot
\left.
  (\partial^\mu\psi_j
   +\frac12(\varphi\partial^\mu A_j-A_j\partial^\mu\varphi)
 +\frac{1}{3\sqrt{3}}\epsilon^{k_2l_2}A_jA_{k_2}\partial^\mu A_{l_2})
\right).
\label{3DL1}
\eeqa

Let us now show how the $G_{2(+2)}/SO(4)$ nonlinear sigma model is
made up of these scalar fields. The adjoint representation of $G_2$ is
decomposed into the direct sum of representations of its subalgebra
$SL(3)$ as ${\bf 14}={\bf 8}\oplus{\bf 3}\oplus{\bf \overline{3}}$.
Let $V$ and $V^*$ be the representation spaces of the ${\bf 3}$ and
${\bf \overline{3}}$. It is known~\cite{Fr} that if we define the Lie 
brackets among $X_{i'}^{~j'},Y_{i'}^{~j'}\in SL(3,\CC)$,
$v_{i'},w_{i'}\in V$ and $v^{*i'},w^{*i'}\in V^*$ by
\beqa
&&[X, ~~Y]_{i'}^{~j'}
=X_{i'}^{~k'}Y_{k'}^{~j'}-X_{k'}^{~j'}Y_{i'}^{~k'},\n
&&[X,~~v]_{i'}
=X_{i'}^{~l'}v_{l'},\rule{0mm}{7mm}\n
&&[X,~~v^*]^{i'}
=-X_{l'}^{~i'}v^{*l'},\rule{0mm}{7mm}\n
&&[v,~~w]^{i'}
=\frac{2}{\sqrt{3}}\epsilon^{i'j'k'}v_{j'}w_{k'},\n
&&[v^*,~w^*]_{i'}
=\frac{2}{\sqrt{3}}\epsilon_{i'j'k'}v^{*j'}w^{*k'},\n
&&[v,~~w^*]_{i'}^{~j'}
=-(v_{i'}w^{*j'}
 -\frac{1}{3}\delta_{i'}^{j'}v_{k'}w^{*k'}),
\label{Fre}
\eeqa
then they realize the Lie algebra $G_2$. If the relations (\ref{Fre}) 
are regarded as those for a real Lie algebra, they generate $G_{2(+2)}$.

Any element of the Lie algebra $G_{2(+2)}$ can be specified by 
a triple $Y_{i'}^{~k'}\in SL(3,\RR)$, $w_{i'}\in V\simeq\RR^3$, 
$w^{*i'}\in V^*\simeq\RR^3$.
We write out the adjoint representation matrix 
\beqa
\mbox{ad}[Y,w,w^*]\equiv&\left[
\begin{array}{ccc}
Y_{i'}^{~k'}\delta_{l'}^{j'}-Y_{l'}^{j'}\delta_{i'}^{~k'}&
  w^{*j'}\delta^{k'}_{i'}
  -\frac{1}{3}w^{*k'}\delta_{i'}^{j'}&
  -(w_{i'}\delta_{k'}^{j'} 
  -\frac{1}{3}w_{k'}\delta_{i'}^{j'})\\
&&\\
-(w_{l'}\delta_{i'}^{k'}
-\frac{1}{3}w_{i'}\delta_{l'}^{k'})&
Y_{i'}^{~k'}&
\frac{2}{\sqrt{3}}\epsilon_{i'j'k'}w^{*j'}\\
&&\\
~w^{*k'}\delta^{i'}_{l'}
-\frac{1}{3}w^{*i'}\delta^{k'}_{l'}&
\frac{2}{\sqrt{3}}\epsilon^{i'j'k'}w_{j'}&
-Y_{k'}^{~i'}
\end{array}\right],
\label{adYww*}
\eeqa
where the indices are so assigned that this matrix maps a triple 
$[Y_{k'}^{~l'}, w_{k'}, w^{*k'}]$ to 
$[Y_{i'}^{~j'}, w_{i'}, w^{*i'}]$.
In this notation the Chevalley generators are given by
$h_1=\mbox{ad}[H_1,0,0]$,
$e_1=\mbox{ad}[E_1,0,0]$,
$f_1=\mbox{ad}[F_1,0,0]$,
$h_2=\mbox{ad}[-H_1+H_2,0,0]$,
$e_2=\mbox{ad}[0,\sqrt{3}n_2,0]$,
$f_2=\mbox{ad}[0,0,-\sqrt{3}n_2]$,
with
$H_1=\left[\begin{array}{ccc}1&&\\&-1&\\&&0\end{array}\right]$,
$E_1=\left[\begin{array}{ccc}0&1&\\&0&\\&&0\end{array}\right]$,
$F_1=\left[\begin{array}{ccc}0&&\\1&0&\\&&0\end{array}\right]$,
$H_2=\left[\begin{array}{ccc}0&&\\&1&\\&&-1\end{array}\right]$
and $n_2=\left[\begin{array}{c}0\\1\\0\end{array}\right]$,
while the $SO(4)$ subalgebra is spanned by the elements in the form 
$\mbox{ad}[U,u,u]$, $U^{\rm T}=-U$.

To identify the physical fields we define 
$A_{i'}\equiv\delta_{i'}^i A_i$, $\varphi^{i'}\equiv\delta_3^{i'}\varphi$
and ${\cal V}_-\equiv\exp(-\ad[0,A,\varphi])$.
Then one obtains $\partial_{\mu}{\cal V}_-\cdot {\cal V}_-^{-1}
=\ad[Y,w,w^*]$,
where $Y,w,w^*$ are given by 
\beqa
Y_{i'}^{~l'}&=&\frac12
\left(-A_{i'}\partial_{\mu}\varphi^{l'}
+\varphi^{l'}\partial_{\mu}A_{i'}\right)
+\frac{1}{3\sqrt{3}}
\epsilon^{j'k'l'}A_{i'}A_{j'}\partial_{\mu}A_{k'},\n
w_{i'}&=&-\partial_{\mu}A_{i'} \rule{0mm}{7mm},\\
w^{*i'}&=&-\partial_{\mu}\varphi^{i'}
+\frac{1}{\sqrt{3}}
 \epsilon^{i'j'k'}A_{j'}\partial_{\mu}A_{k'}.\nonumber
\eeqa

We now consider the product
${\cal V}\equiv {\cal V}_+{\cal V}_-$
with
\beq
{\cal V}_+=\left[
\begin{array}{ccc}
V_{a'}^{~i'}V^{b'}_{~~j'}&&\\
&V_{a'}^{~i'}&\\
&&V^{a'}_{~~i'}
\end{array}
\right],
~~~~~~
V_{i'}^{a'}=\left[
\begin{array}{cc}
e_i^{~a}& -e^{-1}\psi_i\\
0&e^{-1}
\end{array}
\right].
\eeq
Using the equation $\partial_{\mu}{\cal V}\cdot{\cal V}^{-1}
=\partial_{\mu}{\cal V}_+\cdot{\cal V}_+^{-1}
+{\cal V}_+(\partial_{\mu}{\cal V}_-\cdot{\cal V}_-^{-1}){\cal V}_+^{-1}$,
one finds 
$\partial_{\mu}{\cal V}\cdot{\cal V}^{-1}
=\mbox{ad}[Z,z,z^*]$,
where 
\beqa
Z_{a'}^{~c'}&=&\left[\begin{array}{cc}
\partial_{\mu}e_a^{~i}\cdot e_i^{~c}&
e^{-1}e_a^{~i}(\partial_{\mu}\psi_i+Y_i^{~3}) \\
0&e^{-1}\partial_{\mu}e
\end{array}\right],\label{Z}\\
Y_i^{~3}&=&
\frac12(-A_i\partial_{\mu}\varphi
+\varphi\partial_{\mu}A_i)
+\frac{1}{3\sqrt{3}}
\epsilon^{jk}A_iA_j\partial_{\mu}A_k,\label{Y}\\
z_{a'}&=&-\delta_{a'}^ae_a^{~i}\partial_{\mu}A_i,\label{w}\\
z^{*a'}&=&\delta^{a'}_3
e^{-1}(-\partial_{\mu}\varphi
+\frac{1}{\sqrt{3}}\epsilon^{jk}A_j\partial_{\mu}A_k).\label{w*}
\eeqa
 
We introduce here the involutive automorphism $\tau$ such that 
$\tau(\mbox{\bf H})=+\mbox{\bf H}$,
$\tau(\mbox{\bf K})=-\mbox{\bf K}$ for
$G_{2(+2)}=\mbox{\bf H}(=SO(4))\oplus\mbox{\bf K}$. 
This implies
\beq
\tau(\mbox{ad}[Z,z,z^*])
=\mbox{ad}[-Z^T,z^*,z].
\eeq
Defining ${\cal M}\equiv \tau({\cal V})^{-1}{\cal V}$ as usual 
\cite{EiFo,BMG},
one obtains the $G_{2(+2)}/SO(4)$ nonlinear sigma-model Lagrangian
\beqa
\mbox{Tr}\partial_{\mu}{\cal M}^{-1}\partial^{\mu}{\cal M}
&=&-\mbox{Tr}\left(\partial_{\mu}{\cal V}\cdot{\cal V}^{-1}
-\tau(\partial_{\mu}{\cal V}\cdot{\cal V}^{-1})
\right)^2\n
&=&-8{\rm Tr}(Z+Z^T)^2-16(z-z^*)^2.
\label{TrdelM^2}
\eeqa
Using the expressions (\ref{Z}) --- ({\ref{w*}),
one finds that eq.~(\ref{TrdelM^2}) is precisely the sigma-model
part of the dualized reduced Lagrangian~(\ref{3DL1}) up to an overall 
constant factor 32. This completes the construction of the 
$G_{2(+2)}/SO(4)$ nonlinear sigma model from the dimensional 
reduction of the $D=5$ supergravity Lagrangian.

\section{Worldsheet structure of the solitonic string}

In this section we study the properties of the magnetic BPS 1-brane 
solution of the $D=5$ supergravity. The metric is given by 
\beq
ds_5^2 = H^{-1} (-dt^2 + dy^2) + H^2 (dr^2 + r^2 d\Omega_2),
\label{metric}
\eeq
where $r$ is the radial coordinate of the three-dimensional
transverse space, $d\Omega_2$ is the area element of the unit sphere 
$S^2$, and $H$ is a harmonic function $1+Q/r$.
The $U(1)$ gauge-field strength has the only non-vanishing components
\beq
F^{ij}=-\sqrt{3} \epsilon^{ijk} H^{-4}\partial_k H
\label{gauge}
\eeq
for the transverse space indices $i,j,k$.
The solution (\ref{metric}), (\ref{gauge}) is a $D=5$ analogue of 
M5-brane~\cite{M5}; it is non-singular and interpolates~\cite{GiTo} the 
two maximally supersymmetric vacua, namely five-dimensional Minkowski space 
and $AdS_3\times S^2$. $D=5$ supergravity also allows a dual electrically  
charged 0-brane solution which may be viewed as an analogue of M2-brane. 
According to the general rule, there is no intersecting 
solution~\cite{A,O}.

The vanishing condition of the supersymmetry transformation of
gravitino gives the Killing spinor equation
\beq
D_M(\omega)\epsilon
+\frac{i}{8\sqrt{3}}(\Gamma_M^{~~PQ}-4\delta^P_M\Gamma^Q)F_{PQ}\epsilon
=0,
\eeq
where $D_M(\omega)\epsilon=(\partial_M-\frac14\omega_{MAB}\Gamma^{AB})
\epsilon$ (since we changed the ``mostly negative'' metric adopted in  
ref.\cite{ChNi} to the ``mostly positive'' one).
In the above soliton background, this equation is reduced to
\beq
(-\partial_a H \cdot\Gamma^{a0} 
-\frac{i}2 \epsilon^{abc}\partial_c H \cdot\Gamma_{0ab})\epsilon =0.
\label{ki}
\eeq
Taking the gamma matrices as $\Gamma^0=i\sigma_2\otimes {\bf 1}$,
$\Gamma^1=-\sigma_1\otimes {\bf 1}$ and
$\Gamma^a=\sigma_3\otimes \sigma_{a-1}$ $(a=2,3,4)$,
the condition~(\ref{ki}) becomes
\beq
\left[\begin{array}{cc}
0&\sigma_{a-1}\partial_a H \\ 0 & 0 \end{array}
\right]\epsilon =0.
\label{ki2}
\eeq
If one decomposes the Dirac spinor $\epsilon$ as
$\left[\begin{array}{c}\epsilon_-\\\epsilon_+\end{array}\right]$,
eq.~(\ref{ki2}) shows that supersymmetry generated by
$\epsilon_-$ is unbroken. Since the worldsheet chirality is defined
by $\Gamma^0\Gamma^1=-\sigma_3\otimes {\bf 1}$, $\epsilon_{\pm}$ have
definite chirality. Thus two-dimensional (4,4) supersymmetry 
is spontaneously broken to (4,0) by the string.

Let us now find the zero modes of this solitonic string. The obvious
ones are the three bosonic zero modes associated with the broken
translational invariance along the transverse directions due to the
presence of the string. There are also (0,4) fermionic zero modes
coming from broken supersymmetry.

It is known that spontaneously, partially broken supersymmetry 
implies worldsheet supersymmetry~\cite{HP}, which in our case 
requires one more right-moving scalar
to match the numbers of bosons and fermions. The missing bosonic
zero mode is a $D=5$ analogue of the self-dual two-form of the
M5-brane~\cite{CHS,KaMi}. To identify this, we consider a small 
fluctuation of the gauge field $\delta A_M$. It satisfies
\beq
\partial_M(E^{(5)}\delta F^{MN})-\frac1{2\sqrt{3}}
\epsilon^{PQRMN}F_{PQ}\delta F_{RM}=0,
\label{flu}
\eeq
where $\delta F_{MN}=2\partial_{[M}\delta A_{N]}$.
We solve this equation by assuming
$\delta F_{MN}=-\delta F_{NM}=\partial_\sigma X \partial_i(H^2)$
for $M=\sigma$ and $N=i$ for some real scalar $X$ which depends 
only on the worldsheet coordinates $\sigma (\equiv t,y)$, and
$\delta F_{MN}=0$ otherwise. In the background~(\ref{metric}) and
(\ref{gauge}), the $N=i$ component of eq.~(\ref{flu}) shows that $X$ is
a free field, and the $N=\sigma$ equation tells us that $X$ must satisfy
\beq
-\eta^{\sigma\tau}\partial_{\tau}X
+\epsilon^{\sigma\tau}\partial_{\tau}X=0~~~~~(\sigma,\tau=t,y),
\eeq
that is, $X$ is a chiral boson.
Thus we have found four bosonic
and four fermionic zero modes in the right-moving sector, and three
bosonic zero modes in the left-moving sector. They should be compared
with the $(0,2)$ tensor multiplet of the M5-brane. 

It is a common feature in string/M theories that a fundamental object 
in one theory appears as a soliton in its dual theory~\cite{HS,D}. 
Therefore it would be interesting to explore the relation of 
this solitonic solution to a certain chiral string theory. 
The dimensions of the target space of this dual string will be six, 
since the four right-moving bosons are expected to arise as the fields 
in the static gauge~\cite{HP,HS}, in which extra two dimensions are 
supplied from the worldsheet coordinates.

The existence of chiral zero modes implies the non-conservation of the
energy-momentum tensor on the string worldsheet~\cite{DLM}. Again as in
the M5-brane case, it can be understood as the energy-momentum inflow
from the bulk. For such a mechanism to work, the three-form gauge
field of M theory requires a gravitational Chern-Simons coupling,
which was indeed verified to exist~\cite{DLM,VW}. In our case the
gravitational anomaly on the string worldsheet is canceled by the
contribution from the bulk if the $U(1)$ gauge field has the coupling\footnote{
This form of correction was discussed in the Calabi-Yau compactification
of M theory~\cite{Fe}.
}
\beq
{\cal L}_{CS}\propto A\wedge {\rm tr}R^2. 
\eeq
The anomaly polynomial~\cite{AGW} for the four Majorana-Weyl spinors is
$4\cdot(-\frac1{48}p_1)=\frac1{24}{\rm tr}R^2$, while the contribution
from the chiral scalar is $1\cdot(-\frac1{24}p_1)=\frac1{48}{\rm tr}R^2$,
adding up to $\frac1{16}{\rm tr}R^2$. Thus the gravitational
anomaly of the string worldsheet predicts the coupling
\beq
{\cal L}_{CS}=\frac{1}{32\pi q} A\wedge {\rm tr}R^2
\eeq
in the bulk theory. Here $q$ is magnetic charge per unit area of the 
string $q=4\sqrt{3}\pi Q$. This term gives
rise to a modification of the field equation for the $U(1)$ gauge
field, or equivalently the Bianchi identity for the dualized two-form
gauge field exactly in the same manner as the M theory case~\cite{DLM}.
The charge $q$ is proportional to the string tension. 
Since the dual object of this string is a 0-brane, one should be able to  
verify this prediction by a one-loop calculation in the original $D=5$ 
supergravity theory (without any compactification!), thereby establishing 
the particle-string duality in five dimensions. Work along this line is 
in progress.

\section{Concluding remarks}
We have seen that $D=5$ simple supergravity resembles $D=11$ supergravity 
both in their group theoretical properties and in the structures of their 
classical solutions. In particular the solitonic string in the $D=5$ theory 
is a very similar object to M5-brane. We hope that this simple model will 
be useful to test the recent ideas developed in string/M theory, such 
as normal bundle anomaly and AdS/CFT correspondence. 

\vskip 5mm
We thank T. Kawai and G. Schr\"oder for helpful discussions. 
S. M. is grateful to International Center for Theoretical Physics and 
Max Planck Institute for Gravitational Physics for kind hospitality.

\newcommand{\NP}[1]{Nucl.\ Phys.\ {\bf #1}}
\newcommand{\PL}[1]{Phys.\ Lett.\ {\bf #1}}
\newcommand{\CMP}[1]{Comm.\ Math.\ Phys.\ {\bf #1}}
\newcommand{\PR}[1]{Phys.\ Rev.\ {\bf #1}}
\newcommand{\PRL}[1]{Phys.\ Rev.\ Lett.\ {\bf #1}}
\newcommand{\MPL}[1]{Mod.\ Phys.\ Lett.\ {\bf #1}}

\end{document}